\documentclass[twoside,11pt]{article}

\usepackage[accepted]{melba}
\usepackage{graphicx}
\usepackage{booktabs}
\usepackage{graphicx}
\usepackage{wrapfig}
\usepackage{amsmath}
\usepackage{amsfonts}
\usepackage{amssymb}
\usepackage{array}
\usepackage{adjustbox}
\usepackage{multirow}
\usepackage{colortbl}
\usepackage{caption}
\usepackage{amsmath}
\usepackage{multicol}
\usepackage{mathtools}
\usepackage{array,multirow}
\usepackage{makecell}
\usepackage{tikz}
\usepackage{xparse}

\usepackage{soul}

\usepackage{color,xcolor}

\usepackage{url}
\usepackage{algorithm, algorithmic}
\usepackage[super]{nth}

\makeatletter
\DeclareRobustCommand\onedot{\futurelet\@let@token\@onedot}
\def\@onedot{\ifx\@let@token.\else.\null\fi\xspace}

\makeatother

\newcommand{\D}[0]{\mathcal D}

\NewDocumentCommand\DownArrow{O{2.0ex} O{black}}{%
   \mathrel{\tikz[baseline] \draw [<-, line width=0.5pt, #2] (0,0) -- ++(0,#1);}
}
\NewDocumentCommand\UpArrow{O{2.0ex} O{black}}{%
   \mathrel{\tikz[baseline] \draw [<-, line width=0.5pt, #2] (0,0) -- ++(0,#1);}
}

\definecolor{rowblue}{RGB}{220,230,240}
\definecolor{myorchid}{RGB}{150,10,30}
\definecolor{myblue}{RGB}{10,30,250}
\definecolor{mygreen}{RGB}{10,120,10}

\newcommand{\resub}[1]{{#1}}

\melbaid{2023:017}  
\doi{https://doi.org/10.59275/j.melba.2023-g3f8}
\melbaauthors{Abulnaga et al.}  
\volume{2}
\firstpageno{527}  
\melbayear{2023}  
\datesubmitted{04/2023}  
\datepublished{12/2023}  

\melbaspecialissue{Perinatal, Preterm and Paediatric Image Analysis (PIPPI) 2022}
\melbaspecialissueeditors{Jana Hutter, Roxane Licandro, Andrew Melbourne, Esra Abaci Turk, Daphna Link-Sourani, Christopher Macgowan}

\def\eqref#1{equation~\ref{#1}}

\def\1{\bm{1}}

\DeclareMathAlphabet{\mathsfit}{\encodingdefault}{\sfdefault}{m}{sl}
\SetMathAlphabet{\mathsfit}{bold}{\encodingdefault}{\sfdefault}{bx}{n}

\newcommand{\R}{\mathbb{R}}

\ShortHeadings{Shape-aware Segmentation of the Placenta in BOLD Fetal MRI Time Series}{Abulnaga et al.}

\title{Shape-aware Segmentation of the Placenta in BOLD Fetal MRI Time Series}

\author{\firstname S. Mazdak \surname Abulnaga \email abulnaga@mit.edu \\
\addr CSAIL/EECS, Massachusetts Institute of Technology, Cambridge, MA, USA \\
\addr MGH/HST Martinos Center for Biomedical Imaging, Harvard Medical School, Boston, MA, USA 
\AND
\firstname Neel \surname Dey \email dey@mit.edu \\
\addr CSAIL, Massachusetts Institute of Technology, Cambridge, MA, USA
\AND
\firstname Sean I. \surname Young \email siyoung@mgh.harvard.edu \\
\addr MGH/HST Martinos Center for Biomedical Imaging, Harvard Medical School, Boston, MA, USA \\
\addr CSAIL, Massachusetts Institute of Technology, Cambridge, MA, USA 
\AND
\firstname Eileen \surname Pan \email eileenp@mit.edu \\
\addr CSAIL/EECS, Massachusetts Institute of Technology, Cambridge, MA, USA
\AND
\firstname Katherine I. \surname Hobgood \email khobgood@mit.edu \\
\addr CSAIL, Massachusetts Institute of Technology, Cambridge, MA, USA
\AND
\firstname Clinton J. \surname Wang \email clintonw@csail.mit.edu \\
\addr CSAIL/EECS, Massachusetts Institute of Technology, Cambridge, MA, USA
\AND
\firstname P. Ellen \surname Grant \email Ellen.Grant@childrens.harvard.edu \\
\addr Fetal-Neonatal Neuroimaging and Developmental Science  Center, Boston Children's Hospital, Harvard Medical School, Boston, MA, USA
\AND
\firstname Esra \surname Abaci Turk \email Esra.AbaciTurk@childrens.harvard.edu \\
\addr Fetal-Neonatal Neuroimaging and Developmental Science  Center, Boston Children's Hospital, Harvard Medical School, Boston, MA, USA
\AND
\firstname Polina \surname Golland \email polina@csail.mit.edu \\
\addr CSAIL/EECS, Massachusetts Institute of Technology, Cambridge, MA, USA
}
\begin{document}

\maketitle
\begin{abstract}
Blood oxygen level dependent (BOLD) MRI time series with maternal hyperoxia can assess placental oxygenation and function. Measuring precise BOLD changes in the placenta requires accurate temporal placental segmentation and is confounded by fetal and maternal motion, contractions, and hyperoxia-induced intensity changes. 
Current BOLD placenta segmentation methods warp a manually annotated subject-specific template to the entire time series. 
However, as the placenta is a thin, elongated, and highly non-rigid organ subject to large deformations and obfuscated edges, existing work cannot accurately segment the placental shape, especially near boundaries. In this work, we propose a machine learning segmentation framework for placental BOLD MRI and apply it to segmenting each volume in a time series. We use a placental-boundary weighted loss formulation and perform a comprehensive evaluation across several popular segmentation objectives. Our model is trained and tested on a cohort of $91$ subjects containing healthy fetuses, fetuses with fetal growth restriction, and mothers with high BMI. Biomedically, our model performs reliably in segmenting volumes in both normoxic and hyperoxic points in the BOLD time series. We further find that boundary-weighting increases placental segmentation performance by $8.3$\% and $6.0$\% Dice coefficient for the cross-entropy and signed distance transform objectives, respectively. 

\end{abstract}

\section{Introduction} \label{sec:intro}
\paragraph{Biomedical motivation.} The placenta delivers oxygen and nutrients to support fetal growth. Placental dysfunction causes pregnancy complications that affect fetal development, leading to a critical need to assess placental function \textit{in vivo}.
Blood oxygen level dependent (BOLD) MRI 
images oxygen transport within the placenta~\citep{sorensen2013changes,turk2019placental} and has emerged as a promising tool to non-invasively study placental function. Temporal analysis of BOLD MRI with maternal oxygen administration can identify contractions~\citep{abaciturk2020placenta,sinding2016reduced}, biomarkers of fetal growth restriction~\citep{luo2017vivo,sorensen2015placental}, predict placental age~\citep{pietsch2021applause}, and can study congenital heart disease~\citep{you2020hemodynamic,steinweg2021t2}.

\begin{figure}[t]
    \includegraphics[width=\linewidth]{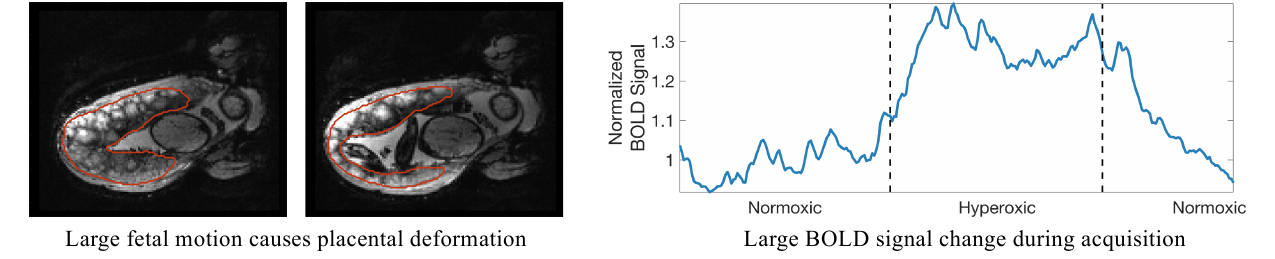}
    \caption{During maternal oxygen administration, the placenta undergoes large shape deformation caused by fetal motion (left) and strong BOLD signal changes caused by maternal hyperoxia (right). Placental boundaries are marked in red. Signal normalization is based on averaging all MRI intensities in the first normoxic period.}
    \label{fig:BOLD-example}
\end{figure}

\paragraph{Challenges and current approaches.} Despite its importance for many downstream clinical research tasks, placental segmentation is often performed manually and can take a significant amount of time, even for a trained expert. For temporal BOLD MRI studies, manual segmentation is rendered more challenging due to the sheer number of MRI scans acquired and rapid signal changes due to common experimental designs. For example, maternal oxygenation experiments acquire several hundred whole-uterus MRI scans to observe signal changes in three stages: normoxia (baseline), hyperoxia, and return to normoxia. During the hyperoxic stage, BOLD signals increase rapidly, 
giving the placenta a hyperintense appearance. Further, placental shape undergoes large deformation caused by maternal breathing, contractions, and fetal movements which are typically stronger during hyperoxia~\citep{you2015robust}, as illustrated in Figure~\ref{fig:BOLD-example}.

Current practice analyzes BOLD signals with respect to a template volume. Deformable registration of all volumes in the time series to the template is performed to enable spatiotemporal analysis~\citep{turk2017spatiotemporal,you2015robust,chi2023dynamic}. However, template-to-volume registration within the uterus can lead to large errors, necessitating outlier detection and possibly rejecting a significant number of volumes~\citep{turk2017spatiotemporal,you2015robust}. These registration difficulties arise from the placenta, fetus, and mother being subject to highly disparate deformation models along the temporal sequence. For example, the fetus undergoes piecewise-rigid motion whereas the placenta deforms highly non-rigidly, thus precluding the direct use of standard registration frameworks such as ANTs~\citep{tustison2020antsx} and VoxelMorph~\citep{balakrishnan2019voxelmorph} on temporal whole-uterus images.

\paragraph{Contributions.} To address these challenges, we propose a deep network framework to automatically segment the placenta in BOLD MRI time series. \resub{Our model is trained on volumes obtained during the normoxic and hyperoxic phases from each patient so as to capture placental shape and appearance variability during maternal oxygenation. }
As the placenta is a thin and elongated organ, we use a boundary-weighted formulation (parameterized by thresholded signed distance function approximations) of several popular region and/or shape-based segmentation objectives which yield significant gains in both placental segmentation accuracy and surface overlap over their non-boundary-weighted equivalents. 
Our model demonstrates consistency in the predicted segmentation label maps on a large dataset of unseen BOLD MRI and 
generalizes to a broad range of gestational ages and pregnancy conditions, thus enabling improved non-invasive pregnancy studies via maternal oxygenation. Finally, to demonstrate the feasibility of our method for clinical research, we illustrate an application based on relative BOLD signal increases.



This paper extends our preliminary analysis of placental segmentation~\citep{abulnaga2022segmentation} first presented at the International Workshop on Perinatal, Preterm and Paediatric Image Analysis held in conjunction with the International Conference on Medical Image Computing and Computer-Assisted Intervention (MICCAI) in 2022. We expand on it by providing more motivation for the work in Section~\ref{sec:intro} and relevant context for boundary-weighted and shape-aware segmentation losses in Section~\ref{sec:related_work}. In our methods in Section~\ref{sec:methods}, we provide additional details, illustrations, and justification for the chosen additive boundary-weighted loss formulation. We then experiment with several additional shape-aware and boundary-weighted loss functions in our experiments (Section~\ref{sec:experiments}) alongside introducing qualitative measures of temporal segmentation performance and providing failure cases. Lastly, we provide a substantially expanded discussion of our work in Section~\ref{sec:discussion} which overviews its place in the shape-aware segmentation literature and its utility for clinical research.

\section{Related Work} \label{sec:related_work}

\paragraph{Placenta segmentation in structural MRI.} Machine learning segmentation models for the placenta have been previously proposed and include both semi-automatic~\citep{wang2015slic} and automatic~\citep{alansary2016fast,torrents2019fully,pietsch2021applause,specktor2021bootstrap} approaches. While semi-automatic methods have achieved success in predicting segmentation label maps with high accuracy, these approaches are infeasible for segmenting BOLD MRI time series due to the large number of volumes that would require manual annotation. The majority of automatic methods focus on segmentation in structural images as opposed to BOLD MRI as in this work. For example,~\citet{alansary2016fast} proposed a model for segmenting T2-weighted (T2w) images based on a 3D CNN followed by a dense CRF for segmentation refinement and validated it on a singleton cohort that included patients with fetal growth restriction (FGR). \citet{torrents2019fully} developed a segmentation framework based on super-resolution and a support vector machine and validated it using a singleton and twin cohort of T2w MRI. \citet{specktor2021bootstrap} focused on transferring segmentation networks across MRI sequences using a self-training model yielding successful segmentation of steady-state precession MRI sequences. For a detailed treatment of fetal MRI segmentation, we refer the reader to the survey by~\citet{torrents2019segmentationsurvey}.

\paragraph{Placenta segmentation in BOLD MRI.} BOLD MR images of the placenta differ greatly from anatomical images, as BOLD images have lower in-plane resolution and the contrast between the placental boundary and surrounding anatomy is less pronounced. Anatomical images may also benefit from super-resolution approaches to increase SNR in the acquired image~\citep{uus2020deformable}. \citet{pietsch2021applause} are the first to consider placental segmentation in functional MRI. They proposed a 2D patch-based U-Net model for functional image segmentation and demonstrated a successful application of age prediction using the estimated T2* values. They focused on a cohort of singleton subjects and demonstrated successful applications to abnormal pregnancy conditions including preeclampsia. In contrast to their approach that segments derived T2* maps, we evaluate our segmentation model on BOLD MRI time series. Furthermore, our 3D model operates on the entire volume rather than patches, thereby helping to better resolve the boundaries of the placenta.

\paragraph{Boundary-weighted segmentation objectives.} Popular supervised segmentation losses such as cross-entropy underperform in the presence of highly imbalanced classes common in radiology volumes. As the interface between organs can be obfuscated by motion-related artifacts and/or isointense appearance, strongly penalizing incorrect predictions near the boundary has been demonstrated to improve segmentation performance across several biomedical applications~\citep{ma2021loss}. For example,~\citet{kervadec2021boundary} and~\citet{zhang2020deep} implemented distance metrics on organ contour predictions in addition to region-based losses such as soft Dice to improve performance. In parallel, several works regress ground truth distance transforms which are strongly weighted near the organ boundary~\citep{hoopes2022synthstrip}. As illustrative examples,~\citet{huang2021shape} regressed ground truth distance maps that are normalized using the Heaviside function to penalize near-boundary misclassifications, while~\citet{karimi2019reducing} proposed a soft-Hausdorff distance loss parameterized by distance transforms. Lastly, the boundary weighting formulation we use is most similar to that of~\citet{caliva2019distance}, wherein a thresholded signed distance transform is used to upweigh boundary neighborhoods in region-based losses. 


\section{Methods} \label{sec:methods}

We train a model $F_{\theta}:\mathbf X\rightarrow \mathbf Y$ parameterized by $\theta$ that takes volumes from a BOLD MRI time series $\mathbf X \in \R^{H \times W \times D \times T}$ and independently predicts a placental segmentation label map $\mathbf Y\in \{0,1\}^{H \times W \times D \times T }$ for each time point $t \in \{1,\dots T\}$, where $T$ is the total number of time points at which MRI scans were acquired. For a given BOLD time series, we have a small number $N_l$ of frames with ground truth labels $(\mathbf x, \mathbf y)$, where $\mathbf x \in \R^{H \times W \times D}$ is an MRI scan and $\mathbf y \in \{0,1\}^{H \times W \times D}$ is the ground truth placenta label map. The model predicts segmentation label maps on each $3$D volume in the time series independently.


\subsection{Architecture and data considerations}
We use a standard 3D U-Net~\citep{ronneberger2015unet} with $4$ blocks in the contracting and expanding paths each. Each block consists of two consecutive \texttt{Conv-BatchNorm-ReLU} blocks using filters of size $3\times 3 \times 3$. Each block is followed by max pooling (contraction path) or transpose convolution (expansion path). We employ batch normalization before \texttt{ReLU} activation. We augment the images using random affine transforms, flips, whole-image brightness shifts, contrast changes, random noise, and elastic deformations, all using TorchIO~\citep{garcia2021torchio}. Specific to segmentation with maternal oxygenation, we simulate the effects of maternal normoxia and hyperoxia with a constant intensity shift in the placenta. All augmentation decisions were made based on cross-validation performance.

To capture the MRI signal and placental shape changes resulting from maternal hyperoxia and fetal motion, we specifically train on several manually segmented volumes in the normoxic or hyperoxic phase. This allows the model to learn from the realistic variations that arise during maternal oxygenation.

\subsection{Additive Boundary Loss}
To emphasize placental boundary details during training, we 
construct an additive boundary-weighting $w_{\delta}$ which is compatible with any per-voxel segmentation loss function $L \left(\cdot \right)$. Given a ground truth placental label map $\mathbf y$, we denote its boundary as  $\partial \mathbf y$. We use a signed distance function $f_{\mathbf y}(x)$ that measures the signed distance, $d(x, \partial \mathbf y)$, of voxel $x \in \R^3$ to the boundary, where  $f_{\mathbf y}(x)>0$ when inside of the placenta and $f_{\mathbf y}(x)<0$ when outside. The boundary weighting is  additive for voxels within $\delta$-distance of $\partial \mathbf y$,

\begin{equation}
\label{eqn-boundary-weighting}
w_{\delta }(x) =
\begin{cases}
w_1 & \text{if } \quad 0 \leq f_{\mathbf y}(x)<\delta, \\
w_2 & \text{if } -\delta<f_{\mathbf y}(x)<0, \\
0 & \text{otherwise}.
\end{cases}
\end{equation}
The weighted loss $L_w$ is then,
\begin{equation}
\label{eqn-boundary-loss}
L_w\left(x \right) = L\left(x \right)\left[w_c\left(x\right)+w_{\delta}\left(x\right)\right],
\end{equation}
where $w_c(x)$ is a per-voxel class weighting.
 In practice, we set $w_2>w_1$, to account for class weighting and to penalize outside voxels more heavily and learn to distinguish the placenta from its surrounding anatomy.
 
\begin{figure}[t]
    \centering
    \includegraphics[width=0.9\textwidth]{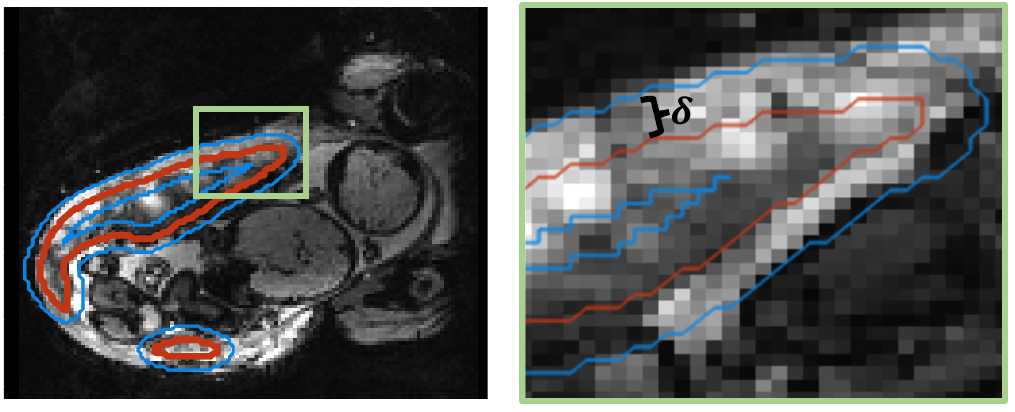}
    \caption{Illustration of inner and outer boundaries used in the additive boundary loss. Right image is zoomed in from the top left corner of the placenta (green box) to closely illustrate boundaries. Placenta segmentation outline is shown in red and inner and outer boundaries $\delta$-distance away are shown in blue.}
    \label{fig:sdf-illustration}
\end{figure}

 We note that several forms of boundary-weighted losses exist in the literature~\citep{ma2021loss} with ours being most similar to~\citet{caliva2019distance}. Many of these use a decaying weight for voxels further from the boundary, while we use a constant weighting.  For computational efficiency, rather than computing $f_{\mathbf y}(x)$ for all voxels $x$ as in~\citet{caliva2019distance}, we approximate the region of distance $\pm \delta$ from $\partial \mathbf y$ using convolutional kernels. To find voxels $x$ with $|f_{\mathbf y}(x)|<\delta$, we estimate a $2\delta$-wide boundary by an average pooling filter on $\mathbf y$ with kernel size $K$ and take the smoothed outputs to lie in the boundary. A larger $K$ produces a wider boundary, penalizing more misclassified voxels. See Figure~\ref{fig:sdf-illustration} for an illustration. Computing the boundary using a convolution operator is advantageous as it does not require computing the signed distance transform directly, which is computationally expensive and bottlenecks deep network training. 

\subsection{Implementation Details}

We train using a learning rate $\eta = 10^{-4}$ with linear decay for $5500$ epochs and select the model with the best Dice score on the validation set. For the additive boundary loss, we set $w_1=1$, $w_2=40$, and $K=11$. \resub{For training, we use a simple preprocessing pipeline. All images are normalized by scaling the \nth{90} percentile intensity value to $1$ without thresholding or clipping any values. We crop or pad all volumes in the dataset to have dimension $112\times112\times80$ and train on the entire 3D volume.  We use a batch size of $8$ MRI volumes in training. } We augment our data with random translations of up to 10 voxels, rotations up to 22$^\circ$, Gaussian noise sampled with $\mu=0,\sigma=0.25$, elastic deformations with $5$ control points and a maximum displacement of $10$ voxels, whole volume intensity shifts up to $\pm 25\%$, and whole-placenta intensity shifts of $\pm0.15$ normalized intensity values. These values were determined by cross-validation on the training set. When evaluating the model on our test set, we post-processed produced label maps by taking the largest connected component to eliminate islands. 
Our segmentation code and trained model are available at \url{https://github.com/mabulnaga/automatic-placenta-segmentation}.

\section{Experiments} \label{sec:experiments}

\subsection{Data}
Our dataset consists of BOLD MRI scans taken from two clinical research studies. Data were collected from $91$ subjects of which $78$ were singleton pregnancies (gestational age (GA) at MRI scan of $23$ weeks (wk), 
 $5$ days (d) to $37$wk$6$d), 
and $13$ were monochorionic-diamniotic (Mo-Di) twins (GA at MRI scan of $27$wk$5$d -- $34$wk$5$d). Of these, $63$ pregnancies were controls, $16$ had fetal growth restriction (FGR), 
and $12$ had high maternal body mass index (BMI, BMI $>30$). Obstetrical ultrasound was used to classify subjects with FGR. For singleton subjects, FGR classification was done based on having fetuses with estimated weight less than the \nth{10} percentile. For twin subjects, FGR classification was determined by monochorionicity and discordance in the estimated fetal weight by growth restriction ($<$\nth{10} percentile) in one or both fetuses; and/or ii) growth discordance ($\geq 20\%$) between fetuses. Table~\ref{tab:subject-demo} shows patient demographics and GA ranges per group.


\begin{table}[t]
\caption{Demographic information for subjects in our analysis. GA is gestational age.}
\label{tab:subject-demo}
    \centering
    \begin{tabular}{ccccc}
    \toprule 
Group & Criteria & Control & FGR & High BMI \\
\midrule
Singleton & \# subjects & $60$ & $6$ & $12$ \\
& GA at MRI & $23$wk$5$d -- $37$wk$6$d & $26$wk$6$d -- $34$wk$5$d & $26$wk$4$d -- $36$wk$6$d \\
\hline
Twin & \# subjects & $3$ & $10$ & $0$ \\
& GA at MRI & $31$wk$2$d -- $34$wk$5$d & $27$wk$5$d -- $34$wk$5$d & N/A \\
\bottomrule
    \end{tabular}
\end{table}

MRI BOLD scans were acquired on a $3$T Siemens Skyra scanner (GRE-EPI, interleaved with $3$mm isotropic voxels, TR $=5.8$--$8$s, TE $=32-47$ ms, FA = $90^{\circ})$. To eliminate intra-volume motion artifacts, we split the acquired interleaved volumes into two separate volumes with spacing $3\times 3 \times 6$mm, then linearly interpolate to  $3 \times 3 \times 3$mm. In our analysis, we only consider one of two split volumes, as the signals are redundant between pairs.
Maternal oxygen supply was alternated during the BOLD acquisition via a nonrebreathing facial mask to have three 10-minute or 5-minute consecutive episodes: 1. normoxia (21\% $O_2$), 2. hyperoxia (100\% $O_2$, 15L/min), and 3. a return to normoxia (21\% $O_2$). 

To generate training data, the placenta was manually segmented by a trained observer. Each BOLD MRI time series had $1$ to $6$ manual segmentations, yielding a total of $176$ ground truth labels. The data was then split into training, validation, and test sets: ($65\%/15\%/20\%$: $63/11/17$ subjects). Data was stratified to have proportional distributions of subjects with singleton and twin pregnancies, then to proportionally distribute healthy controls, subjects with FGR, and subjects with high BMI. Our test set had $15$ singleton pregnancies and $2$ twins. Of the singleton pregnancies, $12$ were healthy controls, $1$ had FGR, and $2$ had high BMI. Both twin subjects had FGR. Our test set had a total of $31$ labeled images, none of which were used before final evaluation.

 Each subject in the training set had up to $N_l=6$ ground truth segmentations in the BOLD time series. To prevent bias in sampling images, we train by randomly sampling $1$ of $N_l$ ground truth segmentations for each subject. The length of one epoch is the number of subjects rather than the number of images. Subject-wise random sampling was used to reduce bias from subjects with more ground truth labels.
 





\subsection{Evaluation}

\paragraph{Performance measures.} We first compare the predicted segmentation label maps to ground truth segmentations. We measure similarity using the Dice score (Dice), the \nth{95}-percentile Hausdorff distance (HD95), and the Average Symmetric Surface Distance (ASSD). To evaluate the feasibility of the produced segmentations for clinical research studying whole-organ signal changes, we evaluate the percentage error in the mean BOLD values between our prediction and the ground truth (BOLD error), defined as $100 \times \vert\hat{b}-b\vert/{b}$, where $b$ and $\hat{b}$ denote the mean BOLD signal in the ground truth and in the predicted segmentation, respectively. As estimating whole-organ BOLD signal changes is an important clinical research task, this metric quantifies the appropriate error caused by using automatic placental segmentations. 


\paragraph{Benchmarked segmentation loss functions.} 
We benchmark several popular loss functions and their boundary-weighted extensions to assess their performance on placental segmentation.
We quantify improved performance using the boundary-weighting approach (Eq.~\ref{eqn-boundary-weighting}) by comparing performance of the cross-entropy loss ($\mathcal{L}_{\mathrm{CE}}$) and the signed distance transform (SDT) loss ($\mathcal{L}_{\mathrm{SDT}}$) with their boundary-weighted counterparts ($\mathcal{L}_{\mathrm{BW-CE}}$, $\mathcal{L}_{\mathrm{BW-SDT}}$). The SDT loss poses segmentation as a regression  problem and predicts the signed distance transformation to the placenta boundary. The loss $\mathcal{L}_{\mathrm{SDT}}$ computes the mean-squared error of the predicted SDT from ground truth~\citep{hoopes2022synthstrip}.
We also benchmark performance on the widely used soft Dice loss ($\mathcal{L}_{\mathrm{Dice}}$)~\citep{Milletari2016VNetFC} and a boundary-weighted Focal loss ($\mathcal{L}_{\mathrm{BW-Focal}}$) with $\alpha=0.5$, $\gamma=2$~\citep{lin2017focal}. As segmentation often benefits from hybrid loss functions~\citep{ma2021loss}, we also evaluate various combinations of losses. Lastly, we evaluate two boundary-focused signed distance-based loss functions from~\citet{huang2021shape} ($\mathcal{L}_{\mathrm{Shape}}$) and from~\cite{karimi2019reducing} ($\mathcal{L}_{\mathrm{HD}}$). 

\paragraph{Consistency with hyperoxia.} We evaluate our model's sensitivity to oxygenation by comparing the accuracy of predictions in normoxia and hyperoxia for subjects with multiple ground truth annotations. We compute segmentation performance using metric $m$ in normoxia  ($m_{\mathrm{normoxia}}(\mathbf y_i,  \hat{\mathbf y_i})$), and in hyperoxia ($m_{\mathrm{hyperoxia}}(\mathbf y_i,  \hat{\mathbf y_i})$), where $m_{\mathrm{normoxia}}(\mathbf y_i,  \hat{\mathbf y_i})$ denotes the similarity using metric $m$ of our predicted segmentation $\hat{\mathbf y_i}$ to the ground truth $\mathbf y_i$ for subject $i$ in normoxia. For the evaluation metric $m$, we use the Dice score, HD$95$, ASSD, and percentage BOLD error. To evaluate consistency with oxygenation, we compute the mean absolute error between segmentation performance in both oxygenation phases  $\left\vert m_{\mathrm{normoxia}}\left(\mathbf y_i, \hat{\mathbf y_i} \right)-m_{\mathrm{hyperoxia}} \left (\mathbf y_i, \hat{\mathbf y_i} \right) \right \vert$. A low error indicates predicted segmentations are consistent with oxygenation changes in the placenta.

\paragraph{Temporal consistency. } We assess the consistency of our predictions by applying our model to all volumes in the BOLD time series of the test set. Since our volumes are acquired interleaved and split into two separate volumes, we apply our model to every second volume in the time series, yielding a mean of $111.7\pm45.3$  volumes per subject. We measure consistency by comparing the Dice score between consecutive volumes. We qualitatively evaluate segmentation performance across the time series and visualize robustness to fetal motion and oxygenation change.


\subsection{Results}
\begin{figure}[t]
    \centering
    \begin{tabular}{cc}
        \includegraphics[width=.46\linewidth]{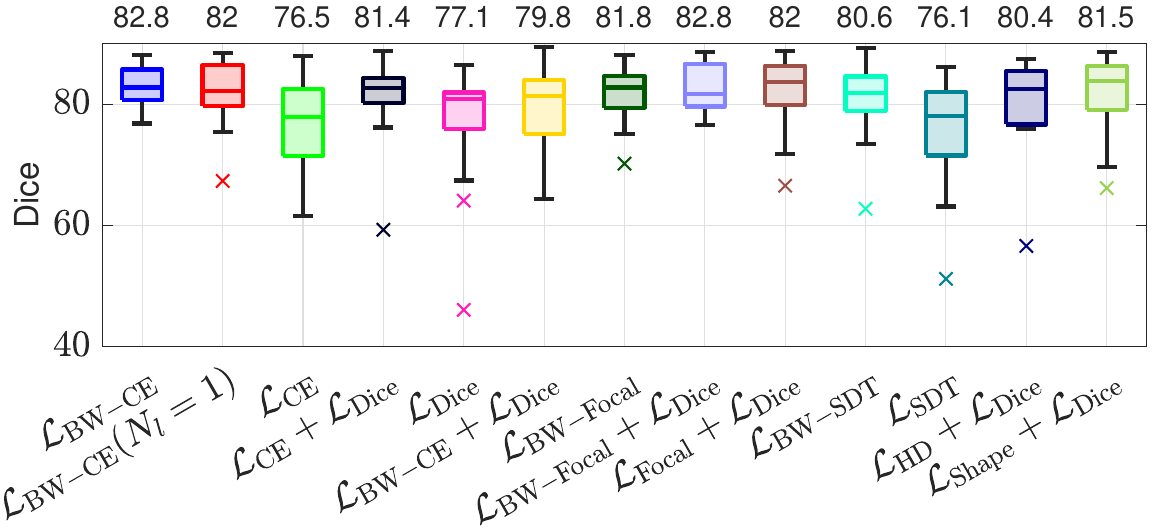} & 
    \includegraphics[width=.46\linewidth]{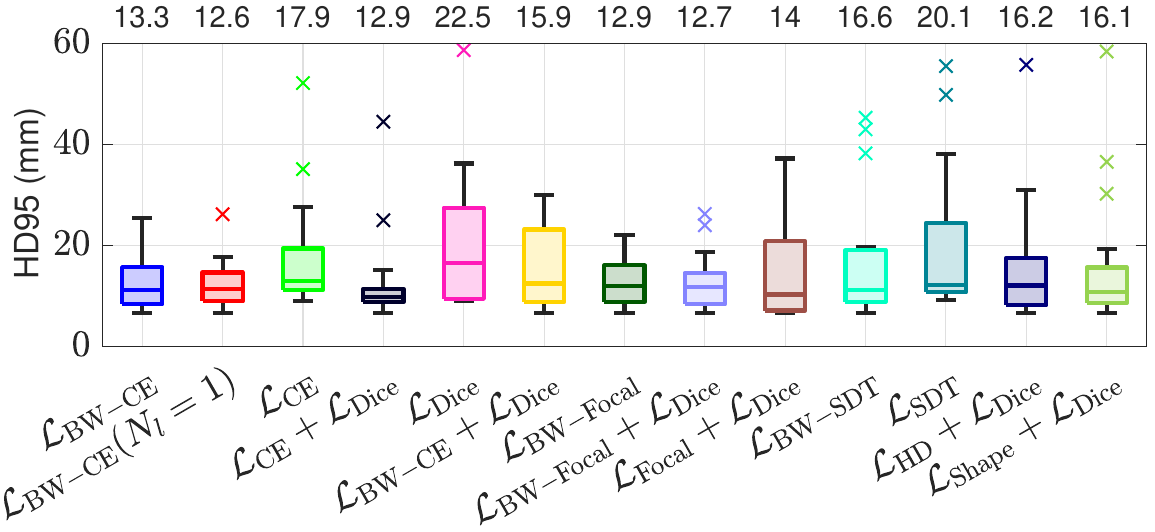} \\
    Dice & HD$95$ \\ \\ 
        \includegraphics[width=.46\linewidth]{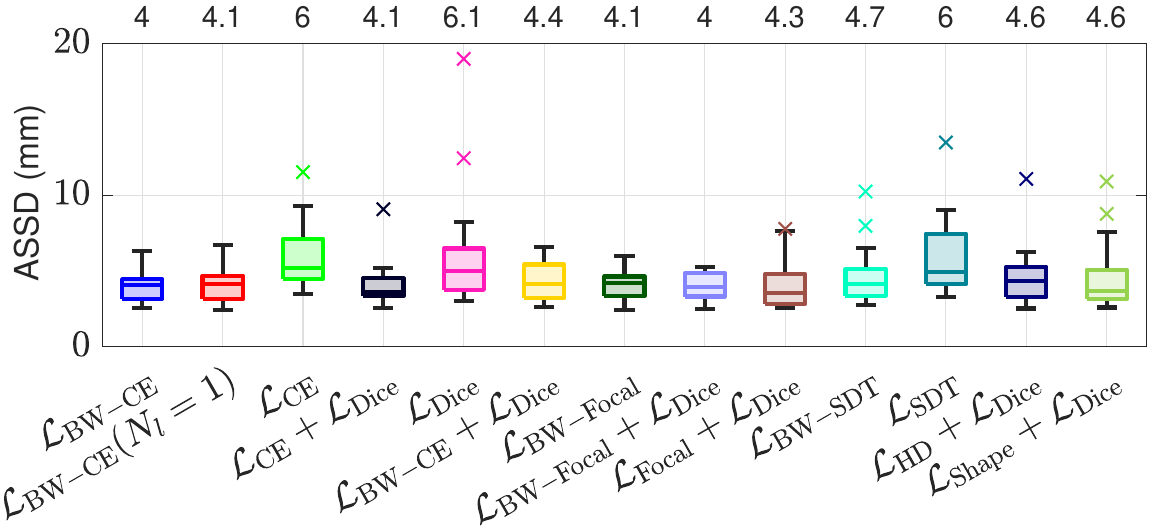}  &
        \includegraphics[width=.46\linewidth]{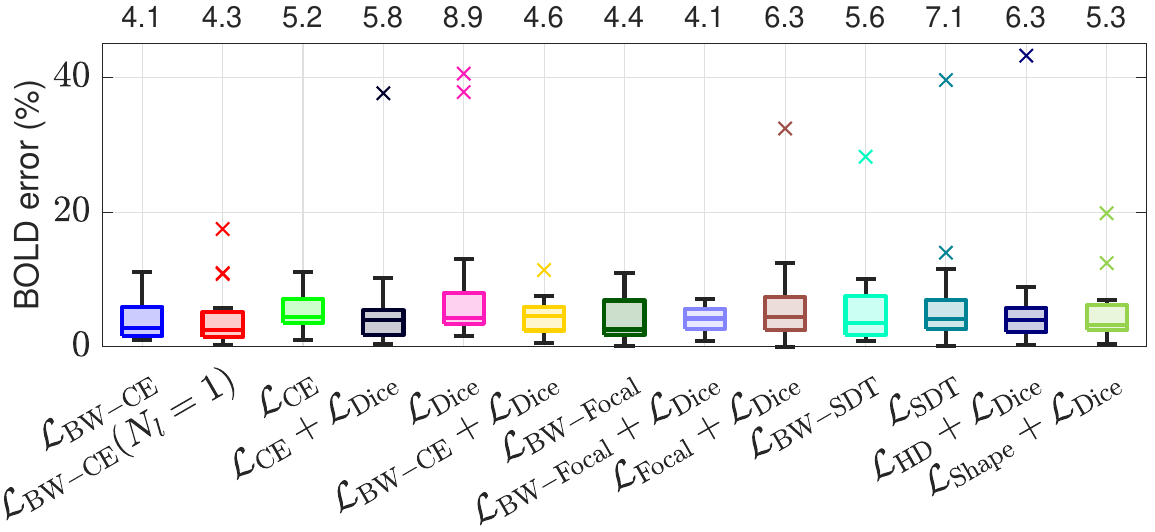}  \\
                ASSD &  BOLD error

    
    \end{tabular}
    \caption{\resub{Box-and-whisker plots of Dice (top left), HD$95$ (top right), ASSD (bottom left), and BOLD error (bottom right) on the test set across all models. We evaluate several commonly used loss functions with and without our boundary weighting and evaluate the baseline shape-based loss ($\mathcal{L}_{\mathrm{Shape}}$) of~\cite{huang2021shape} and the soft-Hausdorff distance loss ($\mathcal{L}_{\mathrm{HD}}$) of~\cite{karimi2019reducing}. Black horizontal lines inside the box indicate the median. The boxes extend to the \nth{25} and \nth{75} percentiles and the whiskers reach to the most extreme values not considered outliers (black bars). The outliers, shown as crosses, are points farther than 1.5
times the interquartile range. Mean values are indicated on the top horizontal axis. Boundary weighting results in higher mean values with more narrow distributions of distortion compared to their non-boundary weighted counterparts. We further observe improved performance compared with baseline losses in terms of mean, variance, and number of outliers. }}
    \label{fig:test-results-whisker}
\end{figure}

Table~\ref{tab:test-results} reports the performance of several segmentation losses on the test set. \resub{Figure~\ref{fig:test-results-whisker} presents box-and-whisker plots of each model's performance.} 
%
Both our best performing models trained using $\mathcal{L}_{BW-CE}$ and $\mathcal{L}_{\mathrm{BW-Focal}}$ achieve a mean Dice score of $82.80$, though $\mathcal{L}_{BW-CE}$ produces slightly lower variance. This model also achieves low relative BOLD error ($4.11\pm 3.0 \%$), indicating that our model's segmentations are suitable for clinical research studies assessing whole-organ signal changes. 
Similar performance is achieved for the other loss functions. Our model also outperforms the two shape-based baselines $\mathcal{L}_{\mathrm{Shape}}+\mathcal{L}_{\mathrm{Dice}}$ and $\mathcal{L}_{\mathrm{HD}}+\mathcal{L}_{\mathrm{Dice}}$ on all metrics, though performance improvements are not statistically significant. \resub{These shape-based baselines demonstrate less consistent performance as they have higher variance with outliers (Figure~\ref{fig:test-results-whisker}).  }

\begin{table}[!t]
\caption{Test results produced by our 3D U-Net model trained using different loss functions. Numbers in bold indicate the best result in each column. We evaluate commonly used loss functions and the shape-based loss ($\mathcal{L}_{\mathrm{Shape}}$) of~\citep{huang2021shape} and the soft-Hausdorff distance loss ($\mathcal{L}_{\mathrm{HD}}$) of~\cite{karimi2019reducing}.}
\label{tab:test-results}
    \centering
    \begin{tabular}{lcccc}
    \toprule 
Loss&Dice&HD$95$ (mm) &ASSD (mm)&BOLD error $(\%)$ \\
\midrule
$\mathcal{L}_{\mathrm{BW-CE}}$&$\mathbf{82.80\pm3.25}$&$13.31\pm6.4$&$\mathbf{4.01\pm1.0}$&$4.11\pm3.0$ \\
$\mathcal{L}_{\mathrm{BW-CE}} (N_l=1)$&$81.98\pm5.3$&$\mathbf{12.61\pm4.66}$&$4.08\pm1.02$&$4.30\pm4.52$ \\
$\mathcal{L}_{\mathrm{CE}}$&$76.47\pm7.43$&$17.92\pm11.02$&$5.96\pm2.1$&$5.22\pm2.53$ \\
$\mathcal{L}_{\mathrm{Dice}}$&$77.13\pm9.74$&$22.51\pm18.02$&$6.14\pm3.95$&$8.95\pm11.37$ \\
$\mathcal{L}_{\mathrm{BW-CE}}+\mathcal{L}_{\mathrm{Dice}}$&$79.82\pm6.57$&$15.91\pm7.93$&$4.44\pm1.3$&$4.61\pm2.59$ \\

$\resub{\mathcal{L}_{\mathrm{CE}}+\mathcal{L}_{\mathrm{Dice}}}$& $\resub{81.44\pm6.42}$ & $\resub{12.87\pm8.9}$ & $\resub{4.14\pm1.42}$ & $\resub{5.84\pm8.28}$ \\

$\mathcal{L}_{\mathrm{BW-Focal}}$&$81.82\pm4.71$&$12.88\pm5.04$&$4.10\pm0.97$&$4.44\pm3.4$ \\
$\mathcal{L}_{\mathrm{BW-Focal}} + \mathcal{L}_{\mathrm{Dice}}$&$82.80\pm3.91$&$12.75\pm5.58$&$4.02\pm0.92$&$\mathbf{4.06\pm1.75}$ \\

$\resub{\mathcal{L}_{\mathrm{Focal}}+\mathcal{L}_{\mathrm{Dice}}}$ & $\resub{81.96\pm6.19}$ & $\resub{14.02\pm8.99}$ & $\resub{4.27\pm1.62}$ & $\resub{6.32\pm7.25}$ \\

$\mathcal{L}_{\mathrm{BW-SDT}}$&$80.63\pm5.96$&$16.63\pm12.4$&$4.68\pm1.92$&$5.64\pm6.35$ \\
$\mathcal{L}_{\mathrm{SDT}}$&$76.06\pm8.67$&$20.14\pm14.25$&$6.01\pm2.51$&$7.09\pm8.86$ \\
$\mathcal{L}_{\mathrm{HD}}+\mathcal{L}_{\mathrm{Dice}}$&$80.39\pm7.16$&$16.16\pm12.17$&$4.62\pm1.98$&$6.35\pm9.52$ \\
$\mathcal{L}_{\mathrm{Shape}} + \mathcal{L}_{\mathrm{Dice}}$&$81.54\pm6.30$&$16.06\pm13.28$&$4.65\pm2.26$&$5.25\pm4.59$ \\

\bottomrule
    \end{tabular}
\end{table}

The boundary weighting improves the performance of several loss functions compared to their non-boundary weighted counterparts. Training the model with our boundary weighting results in a statistically significant increase in performance. When trained using $\mathcal{L}_{\mathrm{CE}}$, we achieve a mean Dice of $82.5$ with boundary weighting in the loss compared to $76.5$ without. Similarly, for $\mathcal{L}_{\mathrm{SDT}}$ we achieve a mean Dice of $80.6$ with boundary weighting compared to $76.1$ without. \resub{However, training with an additive Dice loss improves this performance gap: $\mathcal{L}_{\mathrm{CE}}+\mathcal{L}_{\mathrm{Dice}}$ achieves a Dice of $81.4$ compared to $82.8$ for $\mathcal{L}_{\mathrm{BW-CE}}$, and $\mathcal{L}_{\mathrm{Focal}}+\mathcal{L}_{\mathrm{Dice}}$ achieves a Dice of $82.0$ compared to $82.8$ for $\mathcal{L}_{\mathrm{BW-Focal}}+\mathcal{L}_{\mathrm{Dice}}$. The boundary-weighting also improves mean performance of ASSD and BOLD error and demonstrates more narrow distributions of distortion with fewer outliers (see Figure~\ref{fig:test-results-whisker}.)} Using only the first segmented volume of the BOLD MRI series in normoxia also results in a small drop in performance ($\mathcal{L_{\mathrm{BW-CE}}}$ ($N_l=1$)). Adding labeled examples from the hyperoxic phase helps generalization, as the placental shape and intensity patterns can change greatly.

Our performance is consistent across pregnancy conditions, as we achieve Dice scores of ($77.4$, $88.1$) on the two subjects with twin pregnancies, $82.8\pm3.0$ on the singletons (N$=15$), $83.0\pm5.4$ on the FGR cohort (N=$3$), $82.6\pm3.0$ on the controls (N=$12$) and $(80.6,86.5)$ on the two BMI cases.

\begin{figure}[b]
\scriptsize
    \centering
    \begin{tabular}{ccccc}
        \includegraphics[width=.17\linewidth]{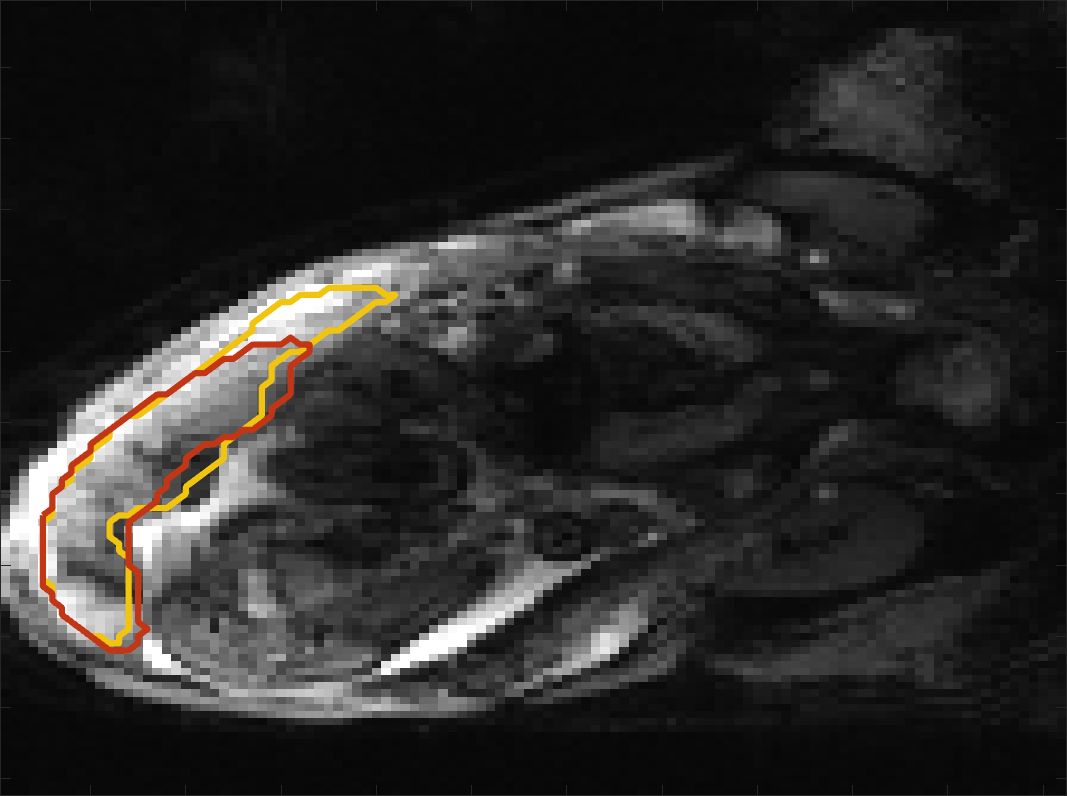} &
    \includegraphics[width=.17\linewidth]{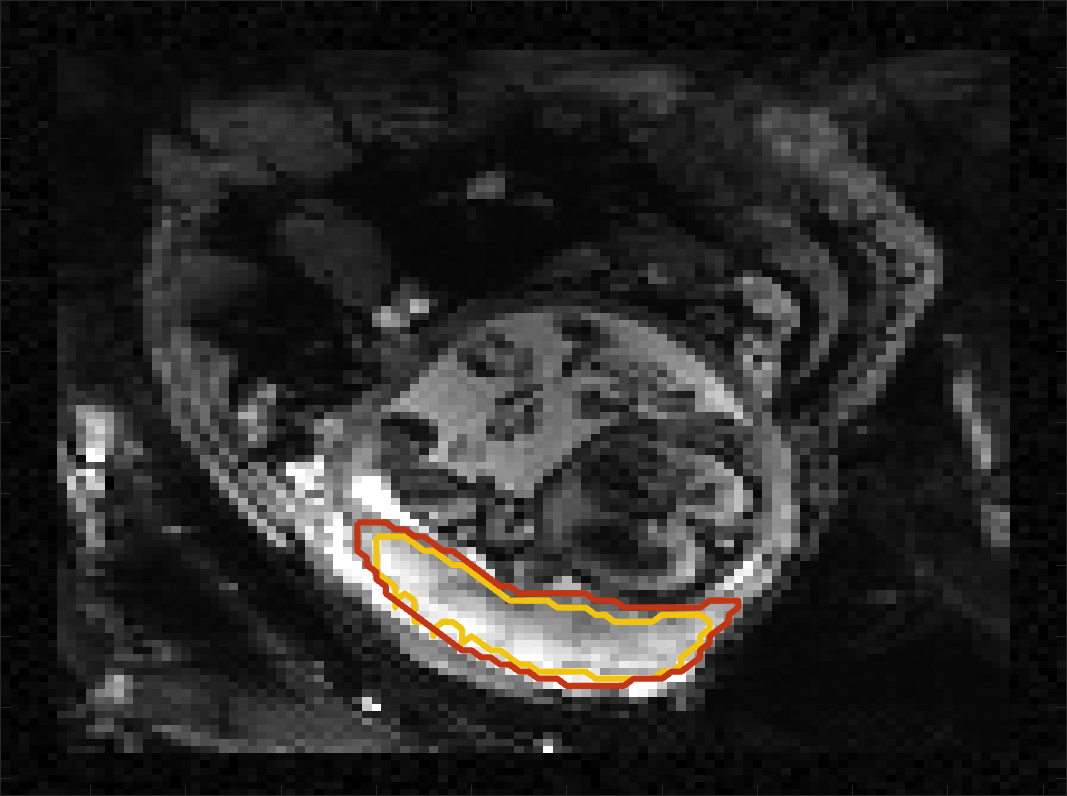} &
        \includegraphics[width=.17\linewidth]{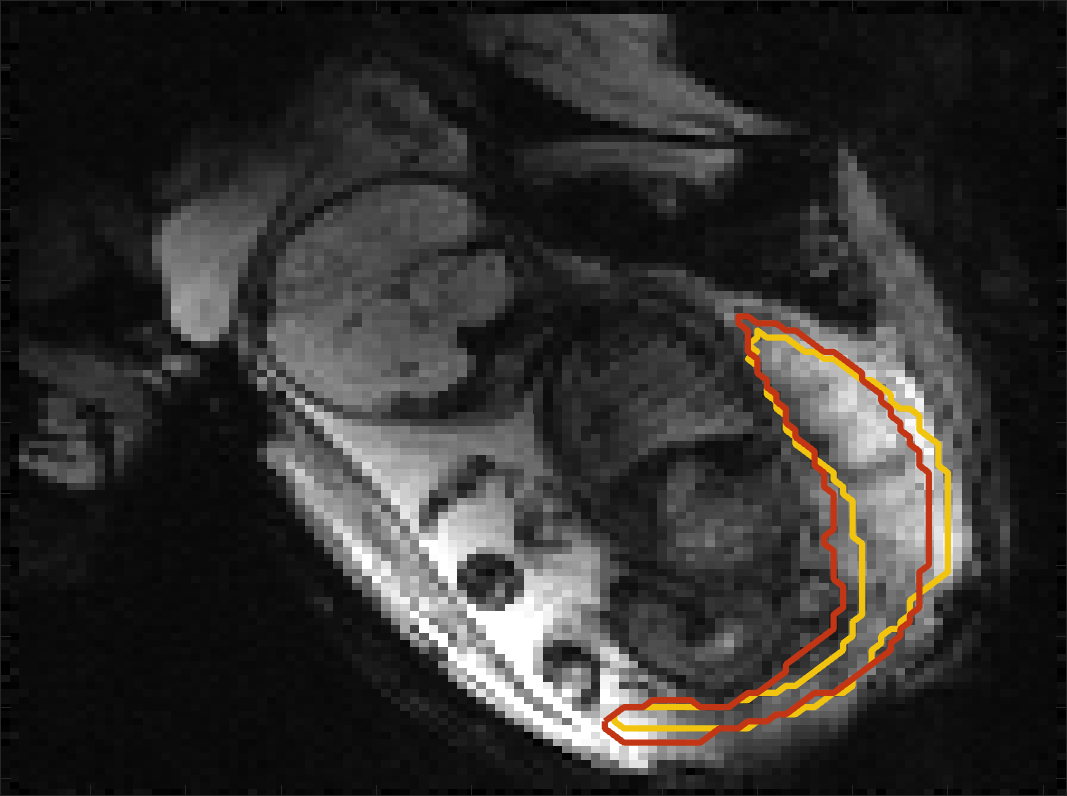} &
        \includegraphics[width=.17\linewidth]{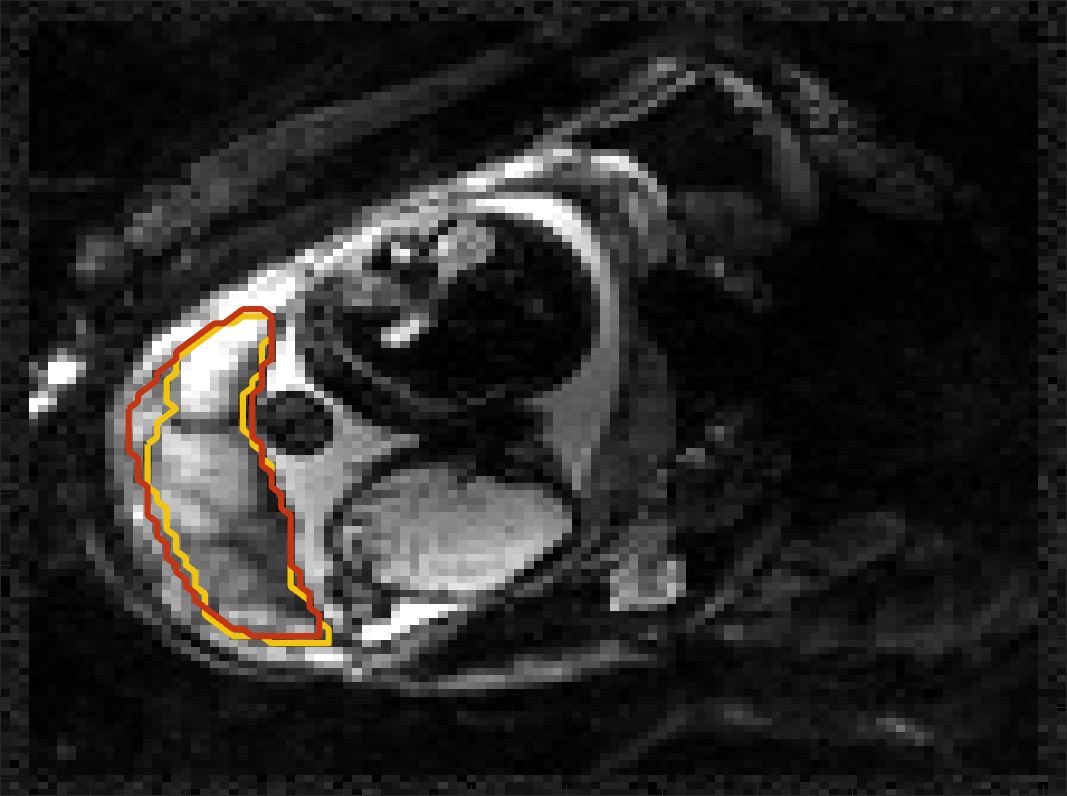} &
    \includegraphics[width=.17\linewidth]{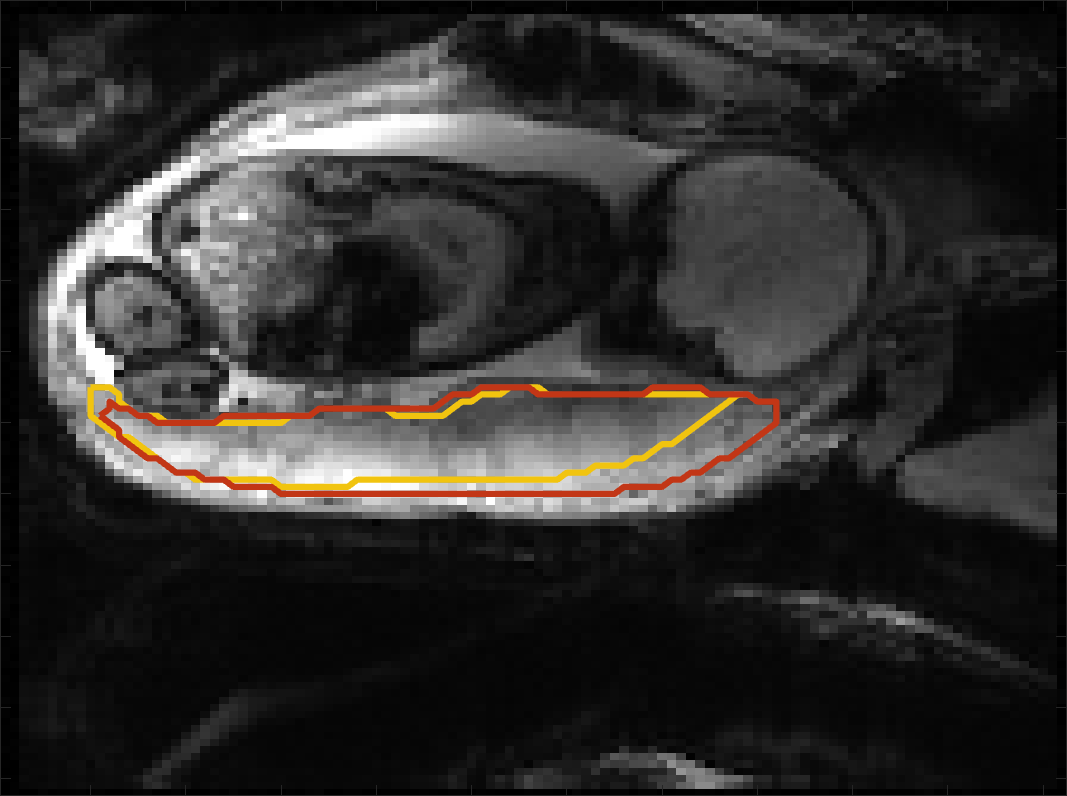} \\ 
    
\includegraphics[width=.17\linewidth]{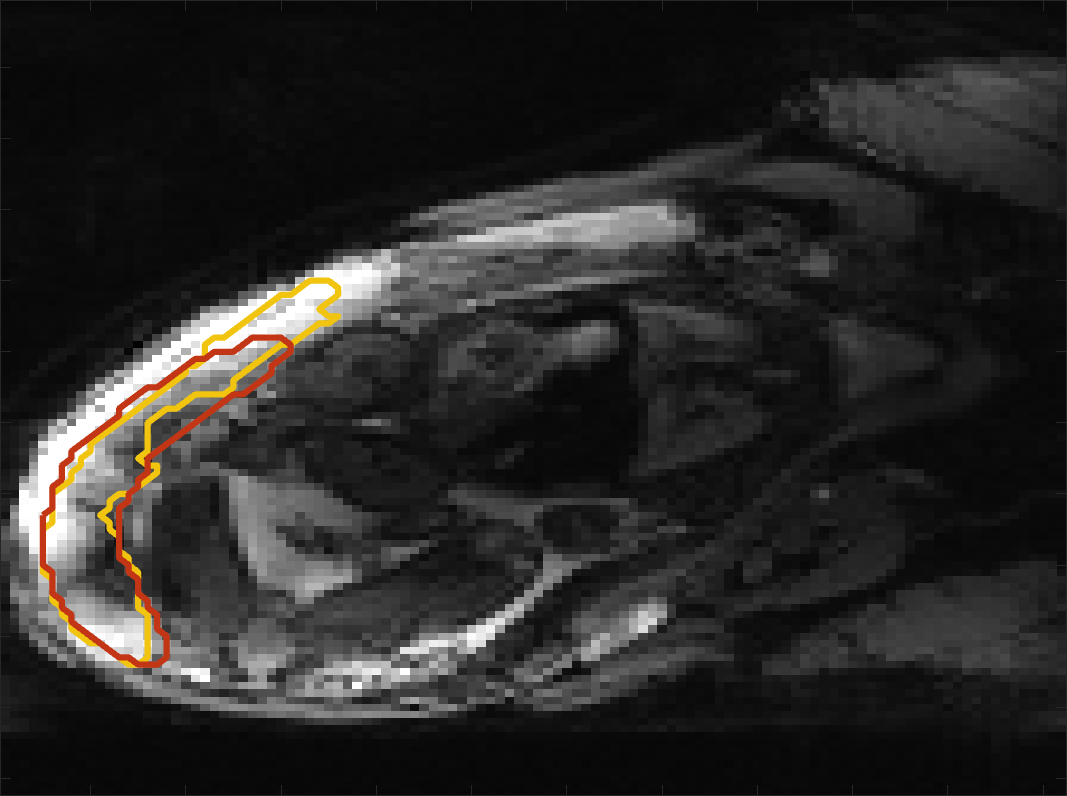}  &
    \includegraphics[width=.17\linewidth]{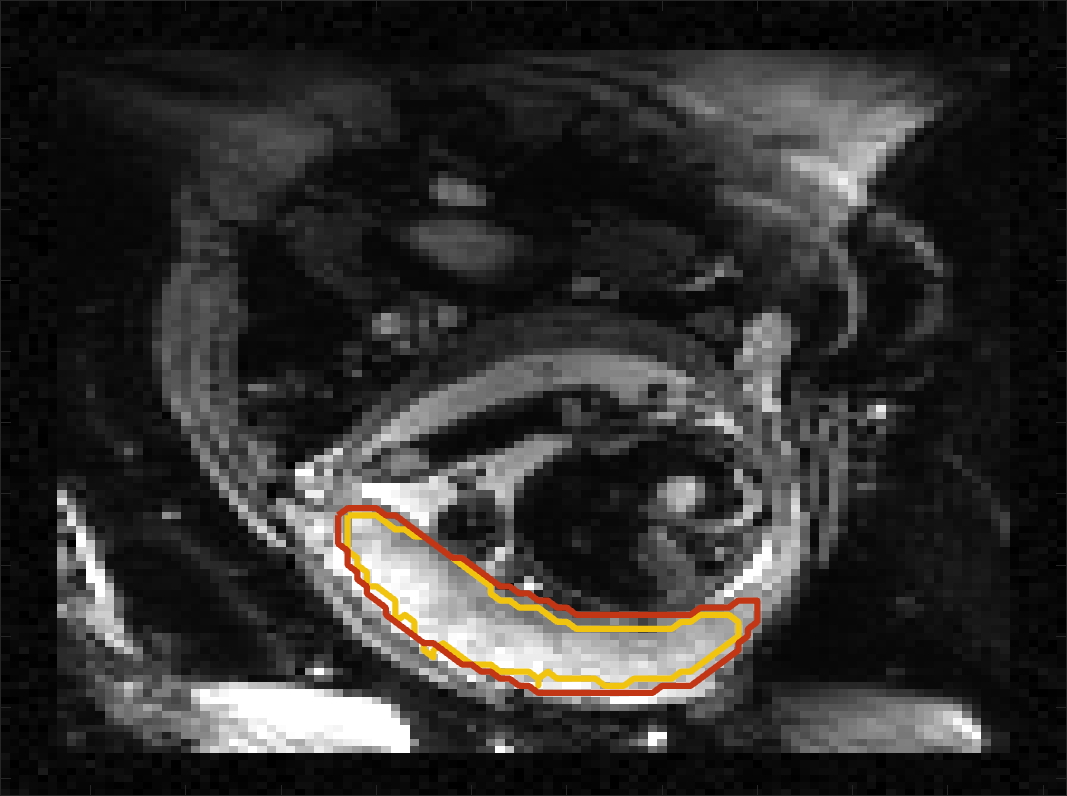} &
        \includegraphics[width=.17\linewidth]{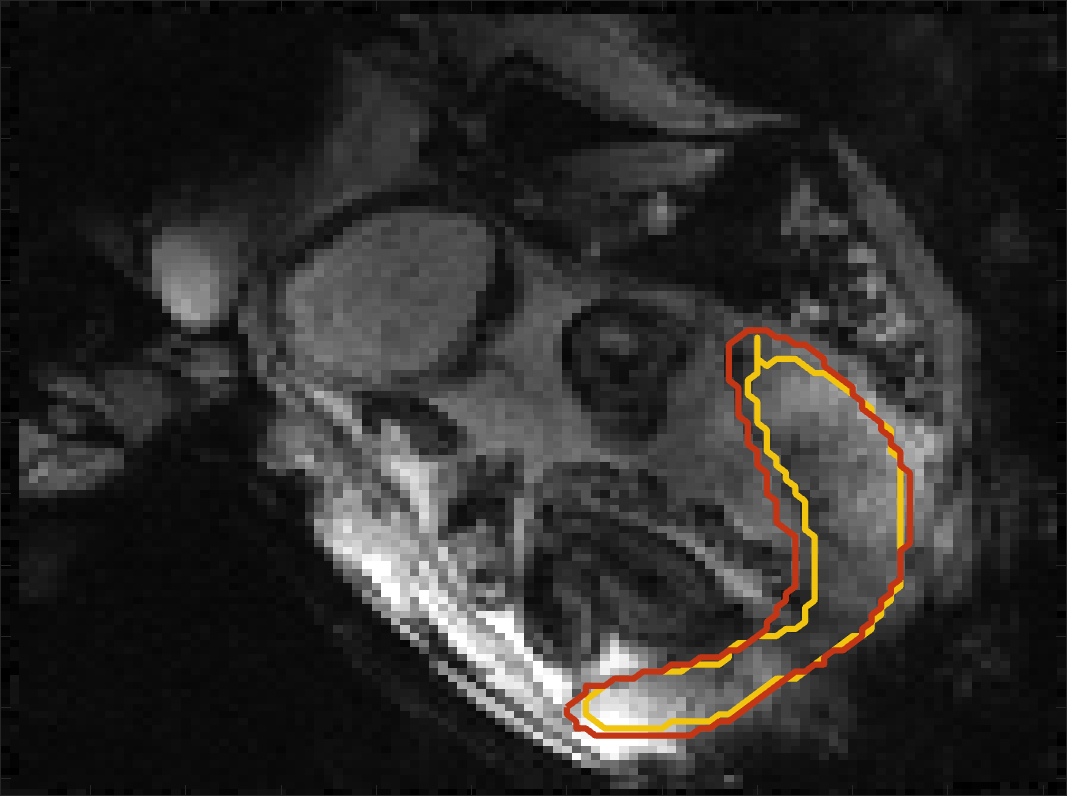} &
    \includegraphics[width=.17\linewidth]{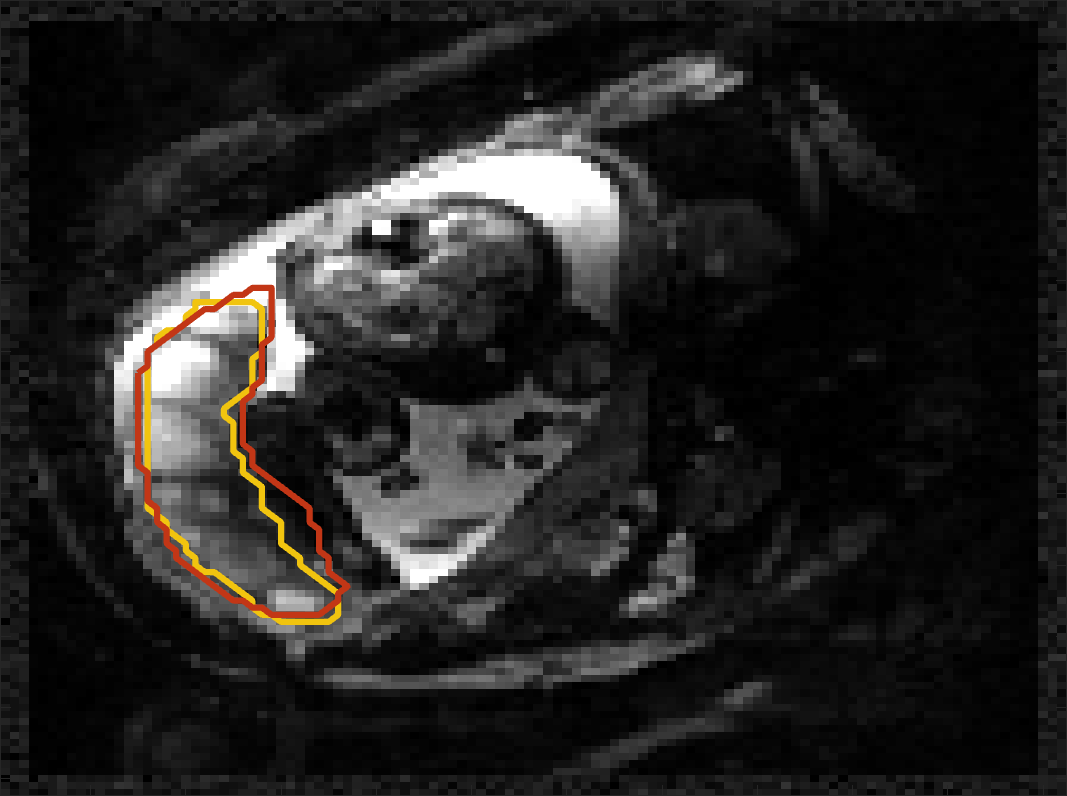} &
        \includegraphics[width=.17\linewidth]{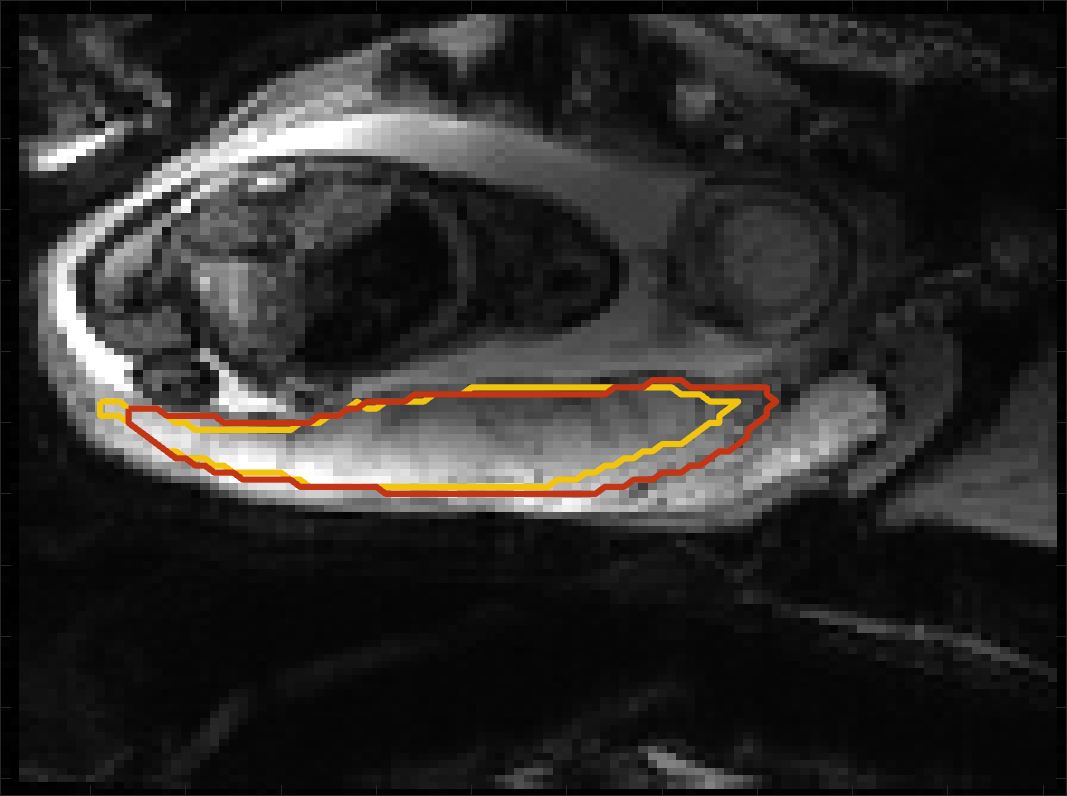} \\
    $74.6$ & $84.4$ & $85.6$ & $86.5$ & $88.0$
    
    
    \end{tabular}
    \caption{Example predictions on $5$ subjects from the test set. Ground truth segmentations are shown in yellow and predictions in red. Dice scores are indicated below each column. Two slices are shown for each subject, spaced $18$mm apart.}
    \label{fig:test-results}
\end{figure}

Our model performs consistently well in the normoxic and hyperoxic phases. For the $5$ subjects with ground truth segmentations in both the normoxia and hyperoxia, we achieve a mean absolute difference between predictions in normoxia and hyperoxia of $3.78\pm1.17$ Dice, $3.15\pm1.90$mm HD95, $0.77\pm0.16$mm ASSD, and $3.21\pm1.36 \%$ relative BOLD error. These results suggest that our model is robust to contrast changes in the placenta resulting from maternal hyperoxia, and can be used in studies quantifying oxygen transport in the organ. A larger number of subjects are needed to assess statistical significance.

Figure~\ref{fig:test-results} compares the predicted label maps with ground truth on $5$ subjects with increasing Dice scores using the BW-CE model. The model accurately identifies the location of the placenta, but in the worst cases misses boundary details. 



\begin{figure}[t]
\scriptsize
    \centering
    \includegraphics[width=\linewidth]{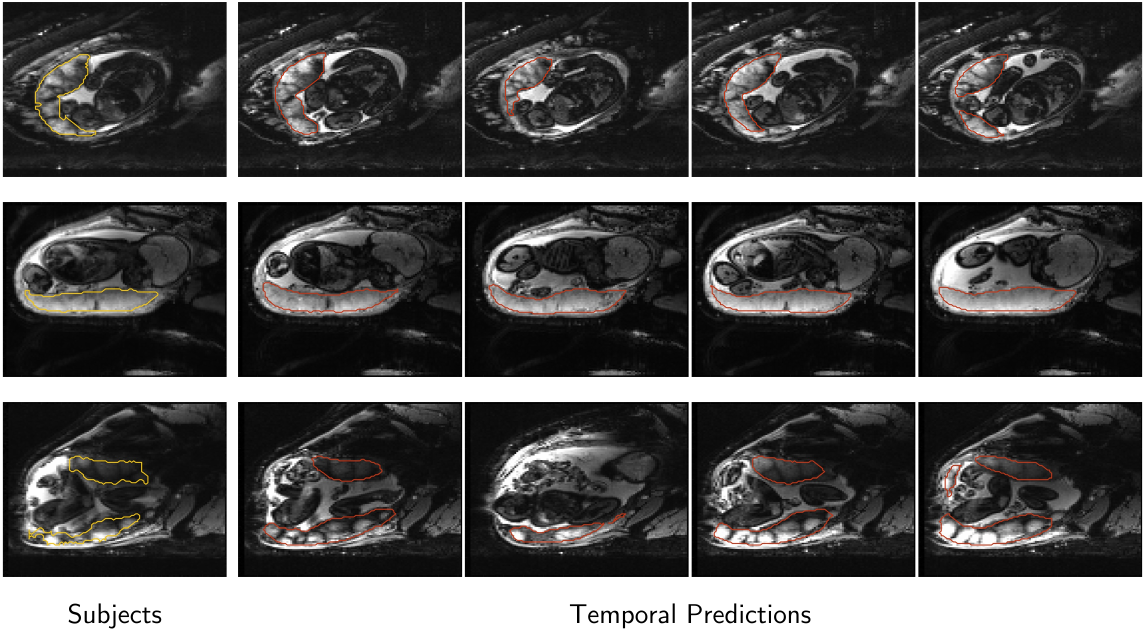}



    \caption{Consistency of predictions throughout the time series on $3$ subjects from the test set. Each row corresponds to a subject. The leftmost column shows the ground truth segmentation outlined in yellow, and temporal-predicted segmentations in columns $2$--$5$ are shown in red. The same axial slice is shown for each subject, and images span the time series. \resub{Segmentations are consistent across time series, achieving consecutive BOLD differences of ($2.6\pm2.0 \%$, $1.1\pm 1.4\%$, and $2.6\pm 2.1\%$) and Dice scores of $90.7\pm0.4$, $95.5\pm0.3$, and $89.7\pm8.8$ (top, middle, and bottom). Variations in these metrics are expected due to fetal motion and changes in oxygenation due to the BOLD protocol.}}
    \label{fig:consistency-time}
\end{figure}

\begin{figure}[t]
    \centering
    \includegraphics[width=0.8\textwidth]{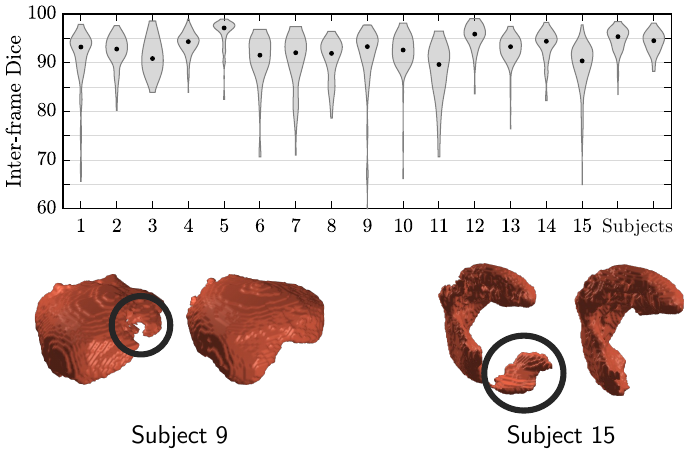}
    \caption{Top: Per-subject density distributions of Dice scores between consecutive predictions in BOLD MRI time series.  Dots inside  distributions indicate the median. Bottom: Example failure cases at frames with Dice $<70$. $3$D models of placentae predicted at two consecutive time points are shown for two subjects. Left: The region in black is distorted in the predicted segmentation of subject $9$. Right: A large portion of the placenta (encircled) is wrongfully predicted in the placenta of subject $15$. Failure cases are likely to occur with large drops in consecutive Dice (Dice $<70$).}
    \label{fig:consecutive-bold}
\end{figure}

\paragraph{BOLD Time Series Evaluation} Figure~\ref{fig:consistency-time} presents example predicted segmentations at multiple points in the BOLD MRI time series for $3$ subjects. The predicted segmentations are robust to large fetal deformations and placental signal changes. Figure~\ref{fig:consecutive-bold} (top) presents distributions of Dice score between predicted label maps of consecutive frames in the BOLD time series for all subjects in the test set. Distributions have high medians (Dice $>90$) for all but one case, with high density at high Dice scores (Dice $>90$). Dice differences are highly affected by fetal and maternal motion that cause placental deformation. We visually verified that modest drops in Dice ($<90$) were mainly due to fetal motion, but $3$ subjects had a small number of frames with large drops (Dice $<70$) that were caused by errors in the produced label maps. Figure~\ref{fig:consecutive-bold} (bottom) shows $3$D models of failed segmentations from two subjects from frames with Dice $<70$. Our model omitted parts of the placenta for Subject $9$ and added a large region for Subject $15$. In practice, these failures ocurred in a small number of frames, $6.8\%$ of frames for Subject $9$ and $0.45\%$ of frames for Subject $15$. Overall, predicted label maps are consistent between consecutive volumes of the MRI time series, achieving a Dice of $92.0 \pm 1.7$ and a BOLD difference of $2.1\pm 0.60 \%$.  The small differences between the relative mean-BOLD values suggest these produced segmentations may be suitable for research studies assessing placental function.

\begin{figure}[t]
\scriptsize
    \centering
    \begin{tabular}{p{0.5\textwidth}p{0.5\textwidth}}
    \includegraphics[width=0.92\linewidth]{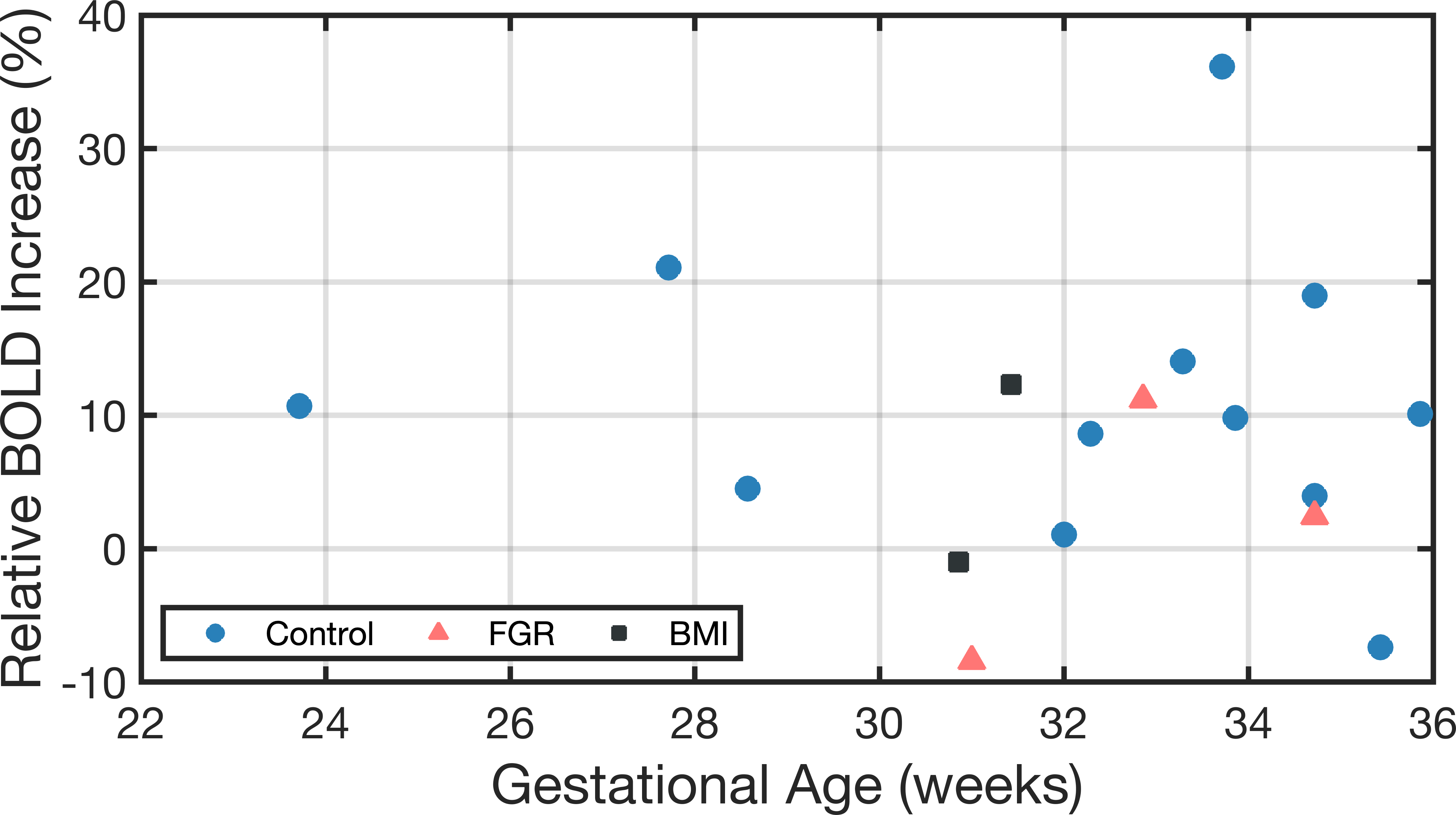} &
    \includegraphics[width=0.92\linewidth]{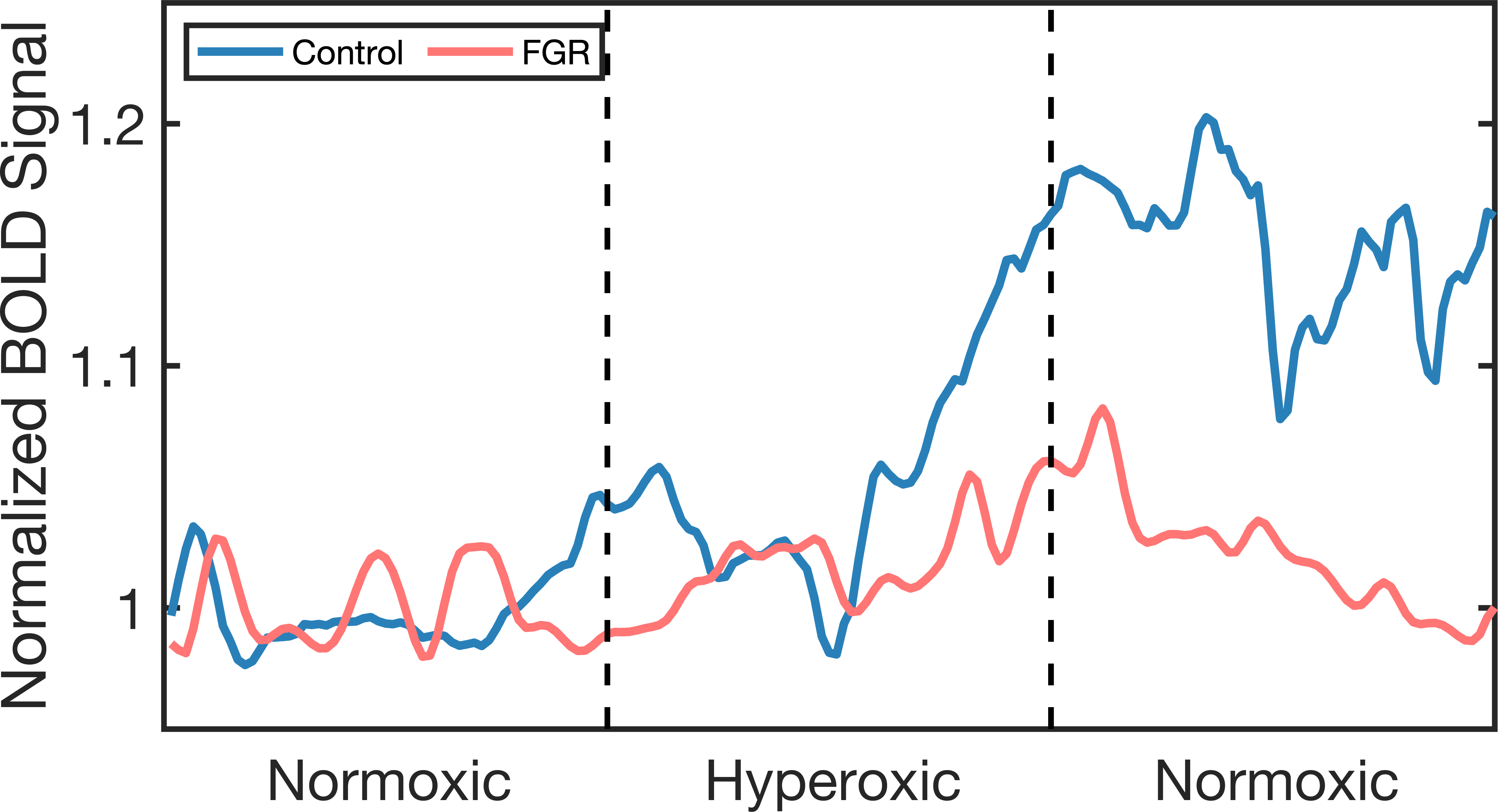} 
    \end{tabular}
    \caption{Example application using our model's produced placenta segmentations in BOLD time series to characterize oxygenation response from maternal hyperoxia. Left: Observed increase relative to normoxic baseline for the test set. Right: Example time series for one singleton control (GA=33wk2d, Dice=$85.5$, $\Delta b=14.1\%$) and one singleton FGR subject (GA=34wk5d, Dice=$83.5$, $\Delta b=2.5\%$).}
    \label{fig:example-application}
\end{figure}

Automatic segmentation of each volume in BOLD MRI time series is advantageous as it can enable whole-organ spatiotemporal analysis without requiring inter-volume registration, which may fail under the presence of large motion. 
We illustrate a possible application of automatic placenta segmentation by investigating the percentage increase in BOLD signal in response to maternal hyperoxia. We calculate the percentage increase over the baseline period: $\Delta b =  \vert b_{H}-b_{N}\vert/b_{N}$, where $b_{N}$ denotes the mean BOLD signal over the baseline period, and $b_{H}$ denotes the mean of the signal in the last $10$ frames of the hyperoxic period. Figure~\ref{fig:example-application} shows a scatter plot of the hyperoxia response for all subjects in the test set and two examples of the BOLD signal time course in the produced placenta segmentation label maps. In the control subjects (N=$12$), we observe an increase of $10.2\pm 11.1\%$. The observed increase for the healthy controls is consistent with previous studies that demonstrated an increase of $12.6\pm5.4\%$ (N=$21$)~\citep{sorensen2015placental} and from $5\%$ to $20\%$ throughout gestation (N=$49$)~\citep{sinding2018placental}. 


\section{Discussion} \label{sec:discussion}

We proposed a model to automatically segment the placenta in BOLD MRI time series and achieved close matching to ground truth labels with consistent performance in predicting segmentations in both the normoxic and hyperoxic phases. Our solution was developed to be resilient to the variability caused by large signal changes in the BOLD experiment protocol. 


\paragraph*{Shape-aware segmentation. } Identification of the placental boundary is challenging as the organ is a thin and elongated structure with limited contrast with surrounding anatomy. 
In this work, we emphasized these challenging aspects of placental shape during network training by using a simple additive boundary-weighted loss function. As hypothesized, boundary weighting significantly improved placenta segmentation performance when integrated with the popular cross-entropy and signed distance transform (SDT) losses as compared with their non-boundary weighted counterparts. 
We then performed an extensive evaluation over shape-based losses including our chosen formulation and the losses of~\citet{karimi2019reducing}, and~\citet{huang2021shape} and found that our adopted loss outperforms others in several key aspects of placenta segmentation.
Lastly, we broadly find that shape-based losses outperform losses without shape information (cross-entropy, Dice, and SDT losses), demonstrating that including shape information can aid in the identification of the placenta.

\paragraph*{Utility for clinical research.} The main objective of this work was to develop a segmentation model to assess whole-organ signal changes in BOLD MRI time series. We achieve low BOLD error (4\%) compared to ground truth with performance that is consistent across oxygenation period. Segmenting each volume in the BOLD MRI time series can be advantageous for clinical research assessing whole-organ temporal changes.
We illustrated one possible study in assessing placental response during hyperoxia and observed an increase in signal intensity consistent with prior work. However, our cohort is limited, and several factors, including maternal position, gestational age, and contractions are covariates not considered. Producing segmentations that are resilient to placental oxygenation can also enable essential post-processing such as motion correction~\citep{turk2017spatiotemporal}, reconstruction~\citep{uus2020deformable}, and mapping to a standardized representation~\citep{miao2017placenta,abulnaga2019placenta,abulnaga2021volumetric,chi2023dynamic}. These tasks are often essential in clinical research studies assessing placental function.

\resub{We provide access to our code and trained model for use in future clinical research studies. Our model is robust to gestational age, pregnancy condition, and oxygenation. Further, the model requires simple preprocessing and can be used with a variety of EPI scans. In this work however, we only trained and tested on isotropic MRI with TE=$32-47$ ms. It is unclear how well the model would work on scans with earlier GA, nonistropic images, or different imaging protocols. }

A limitation of relying only on segmentations is that they can only be used to quantify whole-organ signal changes, such as mean T2$\star$ or mean BOLD increase. One is often interested in assessing functional differences within subregions of the placenta, for example across twins in Mo-Di pregnancies~\citep{luo2017vivo,shnitzer2022automatic}, within vessels~\citep{torrents2019fully}, or across cotyledons~\citep{dey2023anystar}. Localized analysis requires deformable registration~\citep{turk2017spatiotemporal} to track changes within these regions. Having reliable segmentations for each point in the BOLD MRI time series can be used to improve registration, for example by treating segmentations as spatial priors. One may also consider methods that jointly learn registration and segmentation such as that of~\citet{xu2019deepatlas}.

\paragraph*{Comparison to prior work. } The closest work to ours is that of~\cite{pietsch2021applause} that proposes a 2D U-Net based model to automatically segment the placenta in functional MRI (BOLD, T2$\star$). They achieve a Dice of $58$ on a cohort of $108$ subjects of low- and high-risk singleton subjects of a wide GA range. Their performance was comparable to the inter-rater variability of two radiologists (Dice=$68$), which represents an upper limit. In contrast, we achieve a Dice of $82.8$ on a cohort of singleton and twin subjects from healthy pregnancies, subjects with FGR, and subjects with high BMI. While comparing our work with that of~\cite{pietsch2021applause} provides context for our performance, direct comparison with this work is not feasible due to differences in data set size and patient demographics, imaging protocols, and MRI study design. An interesting direction of future work is to quantify the improvement in performance due to model design versus dataset composition. 

\paragraph*{Limitations and Future Work.} The main limitation of this work is the inability to quantify segmentation performance across the entire BOLD MRI study. While we demonstrated low absolute differences in predictions between normoxia and hyperoxia, we only had $5$ subjects with ground truth images in multiple time points. We measured the consistency of consecutive predictions via Dice overlap and percentage BOLD differences. However, without correction for inter-frame placental deformation, the reported scores are subject to noise caused by motion.
We performed visual quality control and found that for many subjects, modest drops in Dice ($<90$), were often due to fetal motion displacing the placenta. In a small number of cases, we observed large drops (Dice $<70$) that were caused by segmentation error (Figure~\ref{fig:consecutive-bold} bottom). Since we apply the model to each volume in the time series independently, imaging artifacts, such as intensity and geometric artifacts, can affect the predicted segmentations.

\resub{We performed a comprehensive evaluation of several commonly used loss functions with and without our boundary weighting approach, and compared with shape-based baselines of two previous works: $\mathcal{L}_{\mathrm{Shape}}$ of~\cite{huang2021shape} and $\mathcal{L}_{\mathrm{HD}}$ of~\cite{karimi2019reducing}. While our boundary-weighting outperformed these loss functions, we observed that any shape-based loss improved performance over conventional loss functions (e.g. $\mathcal{L}_{\mathrm{CE}}$, $\mathcal{L}_{\mathrm{Dice}}$), demonstrating the benefit in capturing the placenta boundary accurately. Additional loss functions exist that we did not compare with, such as the distance transform-based boundary loss of~\citet{kervadec2021boundary}, and the boundary contour-based loss functions of~\citet{specktor2021bootstrap} and~\citet{jurdi2021surprisingly}. Similar to our baselines, these loss functions are additive and aim to improve boundary capture and reduce the Hausdorff distance. Consequently, we do not expect significant differences over our proposed model, though future work should compare with additional baselines. Since our proposed loss is a boundary-based weighting rather than a separate loss function, it is versatile to be used with any existing loss.}

Future work can investigate semi-supervised learning approaches to incorporate all unlabeled volumes in the BOLD MRI time series, increasing the variety of available data to potentially improve temporally consistent segmentation. As there are often a few hundred unlabeled volumes in each BOLD time series, these approaches can more accurately capture the rapid signal changes resulting from fetal motion and maternal oxygenation. The unlabeled data can be incorporated using non-rigid registration as in~\citep{xu2019deepatlas,zhao2019data,chartsias2020disentangle} or by using unsupervised shape-regularization losses~\citep{mirikharaji2018star,young2022sud}.

\section{Conclusion}
We developed a model to automatically segment the placenta in BOLD MRI time series. Our model performed consistently well at different oxygenation phases of the BOLD protocol, and across a variety of pregnancy conditions and gestational ages. We demonstrated one potential clinical research application of this work in quantifying BOLD increase due to hyperoxia that matched reported values from the literature. Automatic segmentation in BOLD MRI time series can be used to investigate oxygenation dynamics in the placenta. For example, temporal segmentations can be used to derive T$2\star$ maps to perform whole-organ signal comparisons across population groups, enabling quantitative analysis of placental function with the ultimate goal of developing biomarkers of placental and fetal health.

\section{Acknowledgments}
This work was supported in part by NIH NIBIB NAC P41EB015902, NIH NICHD R01HD100009, R01EB032708, R21HD106553, MIT-IBM Watson AI Lab, NSERC PGS D, NSF GRFP, and a MathWorks Fellowship.

%
\ethics{Written informed consent was obtained from all subjects. The data used in this work came from two clinical resarch studies. The study protocols were reviewed and approved by Boston Children’s Hospital institutional review board (IRB), from IRB protocol \#P00012416 and \#P00012586. All methods were carried out in accordance with institutional guidelines and regulations.
}


\coi{The authors declare no conflicts of interest.}

\data{The data is currently not approved for public release. The data is being de-anonymized with plans for a future public release. The code and model weights are available at \url{https://github.com/mabulnaga/automatic-placenta-segmentation}.  }




%
%
%
\bibliography{main}
\end{document}